\documentclass[mypaper,7pt,twoside]{CoAst}
\usepackage{epsf,graphicx,fancyhdr}
\renewcommand{\chaptermark}[1]{\markboth{#1}{}}

\renewcommand{\headrulewidth}{0pt}
\newcommand{\kopf}{\small\itshape Comm. in Asteroseismology\\ Vol. 143, 2003}
\newcommand{\Authors}[1]{\begin{center}\normalsize\bf\sf #1 \end{center}}

\renewcommand{\author}[1]{\begin{center}\normalsize\bf\sf #1 \end{center}}
\newcommand{\Address}[1]{\begin{center}\small\sf #1 \end{center}}

\renewenvironment{abstract}{\section*{Abstract}\normalsize\sf}{}
\newcommand{\References}[1]{\begin{flushleft}{\large References\\}\vspace*{2mm}\small #1 \end{flushleft}}

\newcommand{\chapterDSSN}[2]{\chapter[\sf\normalsize #1\\ \footnotesize \hspace*{5mm}by #2 \sf\normalsize][]{#1\\}\rhead[\fancyplain{}{\sf\footnotesize \center{#1}}]{\fancyplain{}{\sffamily\thepage}}\lhead[\fancyplain{\kopf}{\sffamily\thepage}]{\fancyplain{\kopf}{\sf\footnotesize \center{#2}}}}

\renewcommand{\chaptername}{}
\newcommand{\acknowledgments}[1]{\vspace*{5mm}\noindent\begin{bf}Acknowledgments. \end{bf} #1}

\begin{document}
\sf

\chapterDSSN{HD 114839 - An Am star showing both $\delta$ Scuti and
  $\gamma$ Dor pulsations discovered through MOST photometry} {H. King et
al.}

\Authors{H. King$^1$, J.M. Matthews$^1$, J.F. Rowe$^1$,\\
C.Cameron$^1$, R.Kuschnig$^1$, D.B. Guenther$^2$, A.F.J. Moffat$^3$,
S.M. Rucinski$^4$, \\D. Sasselov$^5$, G.A.H. Walker$^1$, W.W.
Weiss$^6$} \Address{$^1$ Department of Physics and Astronomy,
University of British Columbia\\ 6224 Agricultural Road, Vancouver,
British Columbia, Canada, V6T 1Z1\\ $^2$ Department of Astronomy and
Physics, St. Mary's University\\ Halifax, NS B3H 3C3, Canada\\$^3$
D\'{e}partement de physique, Universit\'{e} de Montr\'{e}al\\ C.P.
6128, Succ. Centre-Ville, Montr\'{e}al, QC H3C 3J7, Canada\\$^4$
David Dunlap Observatory, University of Toronto\\ P.O. Box 360,
Richmond Hill, ON L4C 4Y6, Canada\\$^5$ Harvard-Smithsonian Center
for Astrophysics\\ 60 Garden Street, Cambridge, MA 02138, USA\\$^6$
Institut f\"{u}r Astronomie, Universit\"{a}t Wien\\
T\"{u}rkenschanzstrasse 17, A-1180 Wien, Austria}

\noindent
\begin{abstract}
    Using MOST\footnote{MOST is a Canadian Space Agency mission, operated
    jointly by Dynacon, Inc., and the Universities of Toronto and British
    Columbia, with assistance from the University of Vienna.}
    (Microvariability and Oscillations of STars) satellite guide star
    photometry, we have discovered a metallic A star showing hybrid
    p- and g-mode pulsations. HD 114839 was observed nearly continuously
    for 10 days in March, 2005. We identify frequencies in three groups:
    the first centered near 2 cycles/day, in the $\gamma$ Dor pulsation
    range, and two others near 8 and 20, both in the $\delta$ Scuti range.
    This is only the fourth known such hybrid pulsator, including another
    MOST discovery (Rowe et al. 2006, this issue).

\end{abstract}

\section{Introduction}
    Delta Scuti variables are A-F type stars exhibiting both radial and
    non-radial p-mode pulsations with periods ranging from about 1 to 5
    hours.  Gamma Doradus variables are non-radial g-mode pulsators with
    periods ranging from about 7 hours to 3 days (Kaye et al. 1999),
    which overlap the red edge of the $\delta$ Scuti instability strip
    (Handler 1999).

    This region of the HR Diagram was searched by Handler \& Shobbrook
    (2002) for hybrid pulsators exhibiting both p- and g-modes. Their
    search entailed about 270 hr of photometry of a sample of 26 known
    and candidate $\gamma$ Dor stars, to look for $\delta$ Scuti-type
    oscillations.  They discovered one hybrid pulsator, HD 209295, a binary
    system for which the authors argue that excitation of the $\gamma$ Dor
    g-modes is likely due to tidal interaction.  The first known example
    of a single star showing both $\delta$ Scuti and $\gamma$ Dor pulsations
    is the metallic A (Am) star HD 8801 (Henry \& Fekel 2005). The authors
    identified three distinct pairs of frequencies consistent with g- and
    p-modes.

    It was once thought that all Am stars (which do not have strong global
    magnetic fields) had binary companions as the measured values of vsini
    were systematically lower than those for non-peculiar stars in the same
    part of the HR diagram.  The proposed mechanism was tidal breaking in
    short-period ($P \leq 100$ days) binary systems.  Even in the absence
    of a stabilising magnetic field, the slow rotation might reduce
    meridional circulation and other atmospheric turbulence enough to allow
    chemical diffusion to produce the observed abundance anomalies (Abt \&
    Levy 1985). In Abt \& Levy's study of a sample of 55 Am stars, they
    found that 75\% showed evidence of spectroscopic binarity. However,
    the authors estimated that only an additional 8\% of their sample were
    spectroscopic binaries that were missed due to low orbital inclinations,
    suggesting that in fact not all Am stars are in binary systems. Another
    possibility was that slow rotation in single and long-period binary Am
    stars could be explained by evolutionary effects, but more recent studies
    (cf. Henry \& Fekel 2005) suggest that there must be additional factors
    that have yet to be determined.

    Asteroseismology of Am stars could help determine or constrain the
    physical parameters and evolutionary states of these peculiar stars.
    Only a small number of pulsation frequencies has been detected in HD 8801
    (Henry \& Fekel 2005), limiting its potential for seismic modelling.
    To explore the possibility of such modelling, other pulsating Am stars
    with richer eigenspectra must be found.

    We report here on just such a discovery, made with the MOST
    (Microvariability \& Oscillations of STars) space mission (Matthews et al.
    2004, Walker et al. 2003).  The MOST satellite (launched on 30
    June 2003) houses a 15-cm telescope feeding a CCD photometer through a
    custom broadband optical filter.  Its primary mission was to obtain very
    high-precision photometry of bright stars ($V \leq 6$) to detect
    pulsations with amplitudes as low as a few $\mu$mag, with sampling rates
    of better than one exposure per minute and nearly continuous coverage for
    weeks at a time.  MOST's capabilities were improved and extended after
    launch to enable astronomical photometry of the fainter ($11 \leq V \leq
    7$) guide stars in the target fields, used to control spacecraft pointing.

    During March 2005, HD 114839 was one of the guide stars for observations
    of the MOST Primary Science target $\beta$ Comae, and low-amplitude \
    oscillations were evident even in the raw photometry.  HD 114839 had not
    attracted much attention prior to these MOST observations. The only
    accurate stellar parameters for this star available in the literature are
    from the Hipparcos catalogue: $V = 8.46 \pm 0.01$, $B-V = 0.31 \pm 0.01$,
    spectral type = Am, parallax $\pi = 5.04 \pm 1.04$ mas, and $M_V =
    2.06 \pm 0.45$.

\section{Photometry and frequency analysis}

    HD 114839 was monitored nearly continuously for about 10 days during
    22 - 31 March 2005, with only one short gap of about 5 hours during the
    entire run.  The exposure time was 25 sec, sampled every 30 sec.  Our
    final data set contains 27,540 measurements with a duty cycle of 93.5\%.

    As a guide star, HD 114839 was centred on a $45 \times 45$-pixel
    subraster of the MOST Startracker CCD.  The guide star photometry is
    almost entirely preprocessed on board the spacecraft (see Walker et al.
    2005 for additional details).  The mean of the top and bottom rows of
    the subraster is calculated to provide a threshold value of that mean
    plus 30 ADU (Analogue-to-Digital Units).  The intensities of all pixels
    in the subraster above that threshold are summed on board, and this
    value is downloaded to Earth as a flux value for the star, with crude
    sky subtraction.

    The sky background is typically modulated by the 101.4-min orbital
    period of the MOST satellite, due to stray Earthshine (see Reegen et al.
    2005 and Rowe et al. 2006b).  Since no true sky background measurements
    are available for MOST guide star photometry, we subtract from the data
    a running average of the background phased with the MOST orbital period
    (see Rucinski et al. 2004). This automatically suppresses the MOST
    orbital frequency and its harmonics in the Fourier spectrum of the data.

    The frequency analysis for this star was performed with the program
    CAPER (Walker et al. 2005) developed by one of the authors (CC). CAPER
    uses a Discrete Fourier Transform (DFT) to identify frequencies and
    amplitudes, and then obtains a solution via a simultaneous non-linear
    least-squares (Leveberg-Marqardt) fit (Press et al. 1986, p. 678).  We
    refer the reader to Saio et al. (2006) and Cameron et al. (this journal)
    for more details of CAPER.

    The Fourier amplitude spectrum of the HD 114839 photometry is plotted in
    Figure 1, along with the spectral window (which is very clean due to the
    high duty cycle of the 10 days of MOST data).  The initial frequency
    analysis identified 22 significant frequencies, of which 7 can be
    attributed to instrumental or orbital artifacts like the stray light
    modulation mentioned above.  Because the photometry is nondifferential,
    we conservatively reject all power in the DFT below 0.5 cycles/day (c/d).
    Comparison of the Fourier spectrum of the HD 114839 data to the
    DFTs of three other guide stars (fainter by about 1 to 1.5 mag) in
    the same field shows clearly that the peaks we identify at frequencies
    above about 1 c/d are unique and intrinsic to HD 114839.

    The frequencies and amplitudes are listed in Table 1.

    \begin{figure}[ht]
    \begin{center}
    \includegraphics[width=3.5in, height=4.5in,angle=270]{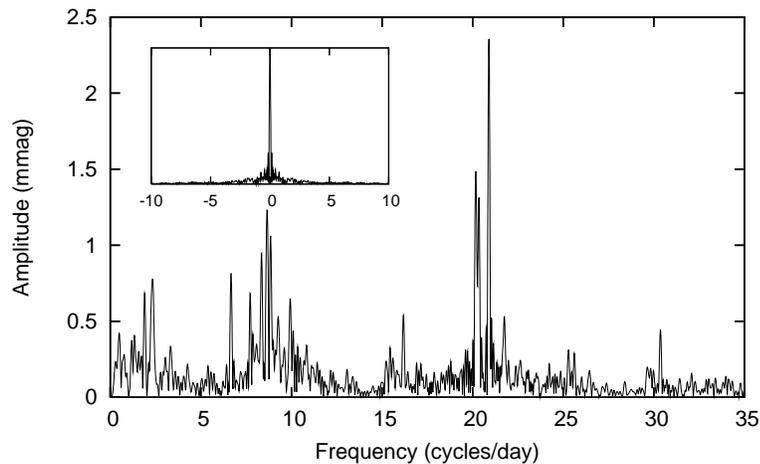}
    \caption{Fourier Spectrum of HD 114839, after removal of known
    instrumental artifacts.  Inset is the window function for the
    highest peak in the spectrum.}
    \end{center}
    \label{Figure 1}
    \end{figure}

    \begin{table}
    \caption{Identified frequencies for the star HD 114839}
    \begin{center}
    \begin{tabular}{|c|c|c|}
    \hline
    Frequency(day$^{-1}$)    &Amplitude(mmag)   &S/N\\
    \hline
    1.3412  &   0.473$\pm$0.004  &   3.87    \\
    1.8905  &   0.637$\pm$0.004  &   4.25    \\
    2.2609  &   0.609$\pm$0.004  &   3.95    \\
    2.3376  &   0.771$\pm$0.004  &   4.75    \\
    6.6678  &   0.794$\pm$0.004  &   4.87    \\
    7.7152  &   0.624$\pm$0.004  &   5.22    \\
    8.3411  &   0.910$\pm$0.004  &   6.94    \\
    8.6477  &   1.210$\pm$0.004  &   8.84    \\
    8.8649  &   1.000$\pm$0.004  &   8.00    \\
    9.2864  &   0.528$\pm$0.004  &   4.32    \\
    9.9378  &   0.593$\pm$0.004  &   4.97    \\
    20.1822 &   1.520$\pm$0.004  &   9.58    \\
    20.3483 &   1.450$\pm$0.004  &   10.02   \\
    20.8976 &   2.260$\pm$0.004  &   15.23   \\
    21.7534 &   0.461$\pm$0.004  &   4.72    \\
    \hline
    \end{tabular}
    \end{center}
    \label{table3}
    \end{table}

\section{A new hybrid pulsator}

    We identify 15 frequencies in HD 114839 with a signal-to-noise (S/N)
    level in amplitude above 3.8.   Four of these frequencies are between
    1 and 2.5 c/d, consistent with $\gamma$ Dor g-mode pulsations, while
    the remaining frequencies are $\delta$ Scuti-type p-modes between
    about 6.5 and 22 c/d.  The p-mode frequencies are grouped into two
    ranges, 7 frequencies around 8 c/d and 4 around 21 c/d.  We note that
    the frequencies found by Henry \& Fekel (2005) in HD 8801 -- another
    Am star -- were clustered in a similar fashion, near 3, 8 and 20 c/d.
    We cannot establish without doubt that HD 114839 is a single star,
    lacking spectroscopic confirmation, but the possible parallels with
    HD 8801 are intriguing.  In any event, this discovery and that by
    Rowe et al. (2006) reported in this issue together double the number
    of known hybrid pulsators, and possibly triple the number of single
    stars which exhibit both $\delta$ Scuti and $\gamma$ Doradus modes
    simultaneously.

\acknowledgments{ JMM, DBG, AFJM, SR, and GAHW are supported by
funding from the Natural Sciences and Engineering Research Council
(NSERC) Canada. RK is funded by the Canadian Space Agency. WWW is
supported by the Austrian Science Promotion Agency (FFG - MOST) and
the Austrian Science Funds (FWF - P17580).
 }

\References{

Abt, H.A, and Levy, S.G. 1985, ApJS, 59, 229\\
Cameron, C. 2006, CoAST, this issue\\
Cox, J.P. 1980, Theory of Stellar Pulsation (Princeton University
Press)\\
ESA 1997, The HIPPARCOS and TYCHO catalogues, Astrometric and
photometric star catalogues derived from the ESA HIPPARCOS Space
Astrometry Mission (ESA Publikations Division, Noordwijk,
Netherlands), ESA SP Ser., 1200\\
Handler, G. 1999, MNRAS, 309, L19\\
Handler, G. and Shobbrook, R.R, 2002, MNRAS, 333, 251\\
Henry, Gregory W. and Fekel, Francis C., 2005, AJ, 129, 2026\\
Kaye, A.B., et al. 1999, PASP, 116, 558\\
Matthews, J.M. et al. 2004, Nature, 430, 51-53\\
Press, W.H. et al. 1986, Numerical Recipes in Fortran 77 (Cambridge
University Press), pg. 678\\
Reegen, P. et al. 2005, MNRAS, 367, 1417\\
Rowe, J.F. et al. 2006a, ApJ, 646, 1241\\
Rowe, J.F. et al. 2006b, CoAST, this issue\\
Rucinski, S.M. et al. 2004, PASP, 116, 1093\\
Saio, H., et al. 2006, astro-ph/0606712\\
Walker, G. A. H., et al. 2003, PASP, 115, 1023\\
Walker, G. A. H., et al. 2005, ApJ, 635, L77\\
}

\end{document}